\documentclass[pre,superscriptaddress,tightenlines,twocolumn,showpacs,longbibliography,floatfix,nofootinbib,notitlepage]{revtex4-1}

\pdfoutput=1

\usepackage{latexsym,amsmath,amsfonts,amssymb,bm,empheq}   
\usepackage{nicefrac} 

\usepackage{multirow}

\usepackage[dvipsnames]{xcolor}
\usepackage{tabularx}   
\usepackage{multirow}

\usepackage{tikz}   
\usepackage{upgreek}

\usepackage{subfigure}  

\definecolor{nblue}{RGB}{28,130,185}
\definecolor{cgreen}{RGB}{76,153,0}
\definecolor{myorange}{RGB}{245,156,74}

\usepackage{hyperref}
\hypersetup{
  colorlinks=true,
  citecolor=magenta,
  urlcolor=-myorange
}

\newcommand{\dd}{\mathrm{d}}
\newcommand{\ddd}{\mathcal{D}}
\newcommand{\ee}{\mathrm{e}}

\newcommand{\ii}{\mathrm{i}}

\makeatletter
\newcommand\thankssymb[1]{\textsuperscript{a}}
\makeatother

\begin{document}

\title{Stochastic dynamics of chemotactic colonies with logistic growth}

\author{Riccardo~\surname{Ben Al\`{i} Zinati}\thankssymb{1}}
\email{riccardo.baz@pm.me}
\affiliation{Sorbonne Universit\'e \& CNRS, \!Laboratoire de Physique Th\'eorique de la Mati\`ere Condens\'ee, LPTMC, F-75005,  Paris, France}

\author{Charlie~\surname{Duclut}\thankssymb{1}}
\email{duclut@pks.mpg.de}
\affiliation{Max-Planck-Institut f\"ur Physik komplexer Systeme, N\"othnitzer Str.~38, D-01187 Dresden, Germany}

\author{Saeed~\surname{Mahdisoltani}\thankssymb{1}}
\email{saeed.mahdisoltani@physics.ox.ac.uk}
\affiliation{Rudolf Peierls Centre for Theoretical Physics, University of Oxford, Oxford OX1 3PU, United Kingdom}
\affiliation{Max Planck Institute for Dynamics and Self-Organization (MPIDS), 37077 G\"ottingen, Germany}

\author{Andrea~\surname{Gambassi}}
\email{gambassi@sissa.it}
\affiliation{SISSA --- International School for Advanced Studies \& INFN, via Bonomea 265, I-34136 Trieste, Italy}

\author{Ramin~\surname{Golestanian}}
\email{ramin.golestanian@ds.mpg.de}
\affiliation{Max Planck Institute for Dynamics and Self-Organization (MPIDS), 37077 G\"ottingen, Germany}
\affiliation{Rudolf Peierls Centre for Theoretical Physics, University of Oxford, Oxford OX1 3PU, United Kingdom}

\renewcommand{\thefootnote}{\alph{footnote}}
\footnotetext[1]{These three authors contributed equally.}
\renewcommand{\thefootnote}{\arabic{footnote}}


\begin{abstract}
The interplay between cellular growth and cell-cell signaling is essential for the aggregation and proliferation of bacterial colonies, as well as for the self-organization of cell tissues.
To investigate this interplay, we focus here on the collective properties of dividing chemotactic cell colonies by studying their long-time and large-scale dynamics through a renormalization group (RG) approach. 
The RG analysis reveals that a relevant but unconventional chemotactic interaction -- corresponding to a polarity-induced mechanism -- is generated by fluctuations at macroscopic scales, even when an underlying mechanism is absent at the microscopic level. This emerges from the interplay of the well-known Keller--Segel (KS) chemotactic nonlinearity and cell birth and death processes.
At one-loop order, we find no stable fixed point of the RG flow equations.
We discuss a connection between the dynamics investigated here and the celebrated Kardar--Parisi--Zhang (KPZ) equation with long-range correlated noise, which points at the existence of a strong-coupling, nonperturbative fixed point.

\end{abstract}

\maketitle

\section{Introduction}

The pervasiveness of emergent self-organization in  living systems shows how local interactions among individuals underpin global structures whose size can significantly exceed the microscopic scale set by each unit.
\textit{Chemotaxis}, namely the ability of cells and synthetic particles to direct their motion according to the spatial gradient of some chemical signals, is one of the primary mechanisms for such self-organization in many biological processes, e.g.\,, immune response and cancer metastasis. 
Phenomenological models describing chemotaxis as a directed motion guided
by the chemical gradients \cite{keller1970initiation,keller1971model} and undergoing stochastic fluctuations \cite{grima2005strongcoupling, golestanian2009anomalous, sengupta2009dynamics, taktikos2012collective} have proven useful in studying chemotactic aggregation (collapse) 
of bacteria~\cite{brenner1998physical, chavanis2004anomalous}, the collective behavior of synthetically active colloids and enzymes  \cite{golestanian2012collective, thakur2012collective,cohen2014emergent,agudo-canalejo2018phoresis,agudo-canalejo2019active},   as well as the interplay between birth and death processes among the living individuals of a colony \cite{gelimson2015collective, kruse2005generic}.

A thorough investigation of the emergent properties of active systems can occasionally be achieved via a bottom-up approach and by coarse-graining the microscopic dynamics of individual particles.
Nevertheless, it is known that such a direct coarse-graining may not necessarily expose all the interaction terms that turn out to be relevant for describing the macroscopic dynamics of the system.
To obtain a sound macroscopic description of the possibly emerging 
collective dynamics, one generally relies on statistical and field-theoretical techniques, such as the renormalization group (RG). These dictate the structure of the relevant interactions that are compatible with the symmetry and conservation laws of the system under investigation~\cite{tauber2014critical,kardar2007statistical}.
Such approaches have been successfully utilized in characterizing the dynamical 
properties of a wide range of biological and synthetic systems~\cite{toner1995longrange,  risler2004universal, caballero2018bulk, cavagna2019dynamicala, Pisegna2021,mahdisoltani2021nonequilibrium,loffler2021behavior,ro2021disorder}. 

In the present work, we investigate the collective behavior of chemotactic particles undergoing birth and death processes by dynamic renormalization group theory. We find that an unconventional chemotactic coupling, previously studied in the context of polarity-induced chemotaxis~\cite{mahdisoltani2021nonequilibrium},  is generated at mesoscopic scales upon coarse-graining the dynamics, even when the underlying polarity mechanism is not explicitly present at the single-particle level.
We therefore extend the analysis presented in Ref.~\cite{gelimson2015collective}, where the polarity-induced chemotactic term was neglected, leading to a theory not closed under renormalization. 
The extended model presents a critical point separating a homogeneous phase with constant density from a collapsed phase characterized by aggregation and nonuniform particle density. 
At that critical point, in the framework of a perturbative dimensional expansion, we find that the RG flow exhibits a runaway behavior, with no real-valued stable RG fixed points in the vicinity of the upper critical dimension.  
However we highlight a connection between our model and the celebrated Kardar--Parisi--Zhang (KPZ) equation with long-range correlated noise \cite{medina1989burgers}, which suggests a possible explanation for the runaway behavior in terms of the existence of a strong-coupling fixed point which is beyond the reach of our analysis.
%

\section{Stochastic chemotactic model with growth}

\subsection{Stochastic Keller--Segel model}

A corpuscular description of the dynamics of the mobile chemotactic particles is obtained by considering $n$ particles moving in a $d$-dimensional space, coupled according to the stochastic Langevin equation
$\dot{\mathbf{r}}_i = \mu_1 \nabla \Phi(\mathbf{r}_i,t) + \bm \xi_i \, ,$
where $\mathbf{r}_i(t)$ describes the position of particle $i$, which undergoes an overdamped motion. 
The trajectory of each particle is biased by gradients of the chemical concentration field $\Phi$ according to the Keller--Segel (KS) model~\cite{keller1971model}; in addition, particles are subject to Gaussian white noises $\bm \xi_i$ with vanishing mean and correlations $\langle \xi_{i,\alpha} (t) \xi_{j,\beta} (t') \rangle = 2D \delta_{ij} \delta_{\alpha\beta} \delta(t-t')$, where $D$ is the noise strength, $i,j=1,\dots,n$ are particle indices, and $\alpha,\beta$ indicate Cartesian coordinates.

As we are interested in the large-scale behavior of the colony, it is convenient to move from a microscopic description, to a coarser one involving a smoothly-varying version $C(\mathbf{r}, t)$ of the particle density $\sum_{i=1}^n \delta^d (\mathbf{r}-\mathbf{r}_i)$.
To include the important fluctuations in the dynamics of the chemotactic colony~\cite{grima2005strongcoupling,newman2004manybody}, we go beyond a mean-field description following the Dean--Kawasaki approach~\cite{dean1996langevin,kawasaki1994stochastic}, which yields a stochastic field theory for the density distribution of chemotactic particles~\cite{chavanis2010stochastic, mahdisoltani2021nonequilibrium}.
The resulting Langevin dynamics describing the colony of chemotactic particles is given by \cite{chavanis2010stochastic}
\begin{align}
    \partial_t C &= D \nabla^2 C -  \mu_1 \nabla \cdot (C \nabla \Phi) + \nabla\cdot \bm \xi^{\rm con} \, . \label{eq:LangevinNoGrowth}
\end{align}
This equation describes the stochastic evolution of the particle density with diffusion coefficient $D$ and with KS chemotactic current $\mu_1 C \nabla \Phi$~\cite{keller1971model}.
The multiplicative noise field $\bm \xi^{\rm con}$, which arises from the individual noises $\bm{\xi}_i$, has a vanishing mean and correlations $\langle \xi^{\rm con}_\alpha (\mathbf{r},t) \xi^{\rm con}_\beta (\mathbf{r}',t') \rangle = 2D C(\mathbf{r},t)\delta_{\alpha\beta}\delta^d (\mathbf{r}-\mathbf{r}')\delta(t-t')$.

We assume that the particles release chemicals at constant rate.
Under the assumption that the diffusion of chemicals is sufficiently fast compared to the particles' dynamics, the chemical field $\Phi$ instantaneously reaches the steady-state profile with $\partial_t \Phi (\mathbf{r},t) \simeq 0$ corresponding to a given $C(\mathbf{r},t)$ and is governed by the screened Poisson equation
\begin{align}
\left(-\nabla^2 + \kappa^2\right) \Phi (\mathbf{r},t) = C (\mathbf{r},t) \, ,
\label{eq_chemical_steadyState}
\end{align}
where $\kappa^{-1}$ is an effective screening length determined by the diffusive spread of the chemical signals before their degradation occurs.

\subsection{Nonlinear birth and death processes}

The combined effect of the cell birth and death processes and their long-range chemotactic interactions are essential to understand the macroscopic properties of growing cell colonies.  
As discussed in Ref.~\cite{gelimson2015collective}, in order to account for birth and decay processes in the coarse-grained description of self-chemotactic systems, we phenomenologically extend the Langevin equation \eqref{eq:LangevinNoGrowth} by considering (i) the cell division process $A\to2A$ with rate $\beta_{\rm d}$, and (ii) the cell 
death/coagulation process $2A\to A$ with rate $\beta_{\rm c}$.
Utilizing a system-size expansion~\cite{vankampen1992stochastic,gillespie2000chemical}, one can encode these microscopic processes into a Langevin description for the particle density within a system of volume $\Omega$, which is valid in the limit of a large number of particles and reads
\begin{align}
    \left. \partial_t C \right|_{\rm growth} = \lambda C(C_0-C) + \xi^{\rm nc} \, .  \label{eq:LangevinGrowthOnly}
\end{align}
Here, the subscript ``growth'' indicates that only the contributions from cell growth and death processes are accounted for in this equation. We have introduced the growth coefficient $\lambda = \beta_{\rm c} \Omega$, the carrying capacity $C_0=(1+\beta_{\rm d}/\beta_{\rm c})/\Omega$, and the nonconserving multiplicative noise $\xi^{\rm nc}$ with correlations
\begin{align}
\begin{split}
    \langle \xi^{\rm nc} (\bm r,t) &\xi^{\rm nc} (\bm r',t') \rangle =\\
    &\lambda C (C_0-2/\Omega+C) \delta^d (\bm r - \bm r')\delta(t-t') \, .
\end{split}
\label{eq_correlationsNonconserving}
\end{align}
Since we are interested in the dynamics in the  asymptotic large-scale and long-time regime, the specific form of the microscopic processes that led to Eq.~\eqref{eq:LangevinGrowthOnly} is irrelevant, provided that $C_0$ is a stable steady state of the growth process.
It is worth mentioning that the Langevin description~\eqref{eq:LangevinGrowthOnly} breaks down close to the absorbing state ($C_0\to0$), as suggested by the fact that, correspondingly, the noise correlations may become negative, leading to a Langevin dynamics with an ill-defined imaginary noise. 
Close to the absorbing state, one thus needs to make use of alternative descriptions of the growth processes that are exact at low densities~\cite{benitez2016Langevin,doering2003interacting}.
In the following, we focus on the case in which the particle density is sufficiently large to be far from the absorbing state and such that the description given by Eq.~\eqref{eq_correlationsNonconserving} is accurate.

\subsection{Dynamics of density fluctuations}

Provided the chemotactic colony is considered sufficiently far from the absorbing state of the dynamics, we expand the  particle density $C$ around the constant carrying capacity $C_0$ as $C(\mathbf{r},t)= C_0 + \rho(\mathbf{r},t)$, where  $\rho\ll C_0$, and subsequently focus on the dynamics of the density-fluctuation field $\rho$. 
The chemical concentration $\Phi$ may also be expanded around a spatially uniform concentration $\Phi_0$ according to
$\Phi(\mathbf{r},t) = \Phi_0 + \phi(\mathbf{r},t)$, where $\phi$ represents the chemical fluctuations that induce chemotactic interactions among the particles~\cite{chavanis2007kinetic}. 
As a result, Eq.~\eqref{eq_chemical_steadyState} relating the chemical concentration and particle density fields can be decomposed as 
$(-\nabla^2 + \kappa^2) \Phi_0 = C_0$  
and
$(-\nabla^2 + \kappa^2) \phi\, = \rho(\mathbf{r},t)$, 
where the first equation admits the uniform solution $\Phi_0 = \kappa^{-2} C_0$. Accordingly, the background chemical concentration $\Phi_0$ is maintained by the uniform part of the particle density~$C_0$, whereas the chemical fluctuation field $\phi(\mathbf{r},t)$ is  induced by the particle fluctuations,  $\rho(\mathbf{r},t)$. 
In the limit in which the chemical signals do not degrade considerably within the system size and are therefore capable of mediating interactions among  distant particles~\cite{chavanis2007kinetic,golestanian2012collective}, the chemical fluctuation $\phi$ is given by the Poisson-like equation
\begin{align}
-\nabla^2\phi(\mathbf{r},t) =  \rho(\mathbf{r},t).
\label{eq_chemical_Poisson}
\end{align}
As we discuss below, focusing on particle and chemical fluctuations $\rho$ and $\phi$ also renders the noise terms in the dynamics additive~\cite{gelimson2015collective, mahdisoltani2021nonequilibrium}.

\subsection{Chemotactic model with growth}

Let us consider the coarse-grained model of chemotaxis that also accounts for birth and decay processes. Combining Eqs.~\eqref{eq:LangevinNoGrowth} and~\eqref{eq:LangevinGrowthOnly}, the stochastic Langevin equation for the density fluctuations $\rho$ reads
\begin{equation}
  (\tau \partial_t - D\nabla^2 + \sigma)\rho = - \lambda \rho^2 - \mu_1 \nabla(\rho\nabla \phi) -\mu_2 \nabla^2 (\nabla \phi)^2 + \zeta \label{eq:Langevin} .
\end{equation}
The linear term $\sigma=  (\lambda -\mu_1) C_0$ introduces a length scale $\xi = \sqrt{D/\sigma}$ over which density fluctuations remain correlated. 
Such correlation length diverges when the control parameter $\sigma$ approaches its critical value $\sigma_c$.
This critical point $\sigma_c$ demarks the transition between a dilute, homogeneous phase for $\sigma>\sigma_c$, and a dense, collapsed one for $\sigma<\sigma_c$~\cite{mahdisoltani2021nonequilibrium}.
In the absence of nonlinearities, the critical value $\sigma_{c,0}=0$
is achieved when the linear part of the chemotactic aggregation term $\mu_1 C_0$ balances the linear decay contribution $C_0 \lambda$. In the presence of nonlinearities, this critical value will be shifted, with a non-universal fluctuation-induced shift that depends on microscopic details.
Note also that we have introduced the time-scale parameter $\tau$ since it eventually acquires a non-trivial renormalization.

In addition to the KS chemotactic nonlinearity $\mu_1 \nabla\cdot(\rho\nabla \phi)$, in Eq.~\eqref{eq:Langevin} 
we have also introduced a second chemotactic nonlinearity $\mu_2 \nabla^2 (\nabla \phi)^2$; 
this unconventional interaction is a relevant coupling in the macroscopic dynamics, and, as it has been shown in Ref.~\cite{mahdisoltani2021nonequilibrium}, it gives rise to novel large-scale phenomena such as superdiffusion and non-Poissonian number fluctuations in the absence of nonlinear growth processes.
The $\mu_2$ coupling may arise from the polarity effects in the microscopic dynamics of chemotactic particles, and therefore this terms is referred to as the \textit{polarity-induced chemotaxis}~\cite{mahdisoltani2021nonequilibrium}.
As it will become clear below, even if this term is not present in the single-particle dynamics of Eq.~\eqref{eq:Langevin}, the polarity-induced coupling is effectively generated in the macroscopic dynamics upon coarse-graining and it is as relevant (in the RG sense) as the KS nonlinear term $\propto \mu_1$ and the nonlinear growth term $\propto \lambda$.
In this respect, the present analysis extends the model of Ref.~\cite{gelimson2015collective} by incorporating the effective polarity-induced coupling in the chemotactic dynamics.

Finally, the noise term $\zeta =  \xi^{\rm nc} + \nabla \cdot \bm \xi^{\rm con}$ in Eq.~\eqref{eq:Langevin} is Gaussian and combines an additive and a multiplicative part.
As long as the system evolves far from the absorbing state, the expansion \mbox{$C=C_0+\rho$} allows us to discard the multiplicative part of the noise $\zeta$, as it is also RG irrelevant~\cite{mahdisoltani2021nonequilibrium}. 
The remaining additive noise has correlations given by 
\begin{equation} \label{eq:zeta-corr-full}
\langle \zeta (\mathbf{r},t) \zeta (\mathbf{r}',t') \rangle = 2(\mathcal{D}_0 -\mathcal{D}_2\nabla^2)\delta^d (\mathbf{r}-\mathbf{r}')\delta(t-t'), 
\end{equation}
where the conserving part $\mathcal{D}_2=D C_0$ comes from the stochastic KS equation \eqref{eq:LangevinNoGrowth}, whereas the nonconserving part $\mathcal{D}_0=\lambda C_0^2$ comes from the expansion\footnote{Note that we have neglected the term $2/\Omega$ compared to $C_0$ in Eq.~\eqref{eq_correlationsNonconserving}, which is consistent with the assumption of being far from an absorbing state and of considering large populations, such that $\Omega\gg 1$.} of the stochastic growth equation~\eqref{eq:LangevinGrowthOnly}.

To conclude this section, let us analyse the symmetries of the stochastic chemotaxis model with growth. The Langevin equation \eqref{eq:Langevin} possesses a trivial $\phi$-shift symmetry and a more subtle pseudo-Galilean symmetry. 
The $\phi$-shift symmetry is a direct consequence of the fact that the dynamics considered here is only sensitive to the gradients of the chemical field $\phi$, which makes the Langevin equation trivially invariant under the shift $\phi \to \phi' = \phi + \text{const}$. 
The pseudo-Galilean symmetry expresses the invariance of the dynamics under the transformation~\cite{mahdisoltani2021nonequilibrium}
\begin{subequations}
\begin{align}
\phi'(\mathbf{r},t) &= \phi( \mathbf{r} + t (\mu_1 - 2\mu_2) \mathbf{v} /\tau, t) - \mathbf{v}\cdot\mathbf{r} \, ,\\
\rho'(\mathbf{r},t) &= \rho( \mathbf{r} + t (\mu_1 - 2\mu_2) \mathbf{v} /\tau, t) \, ,
\end{align}\label{eq:Galilean}%
\end{subequations}%
where $\mathbf{v}$ is an arbitrary $d$-dimensional boost vector. As we discuss below, this symmetry enforces an identity between critical exponents.
Moreover, in the absence of the $\lambda \rho^2$ term, Eq.~\eqref{eq:Langevin} reduces to the number-conserving chemotaxis model considered in Ref.~\cite{mahdisoltani2021nonequilibrium}. 
In that case, the pseudo-Galilean symmetry,  accompanied by the non-renormalization of the noise term, is  such that the dynamical scaling exponents can be determined exactly.
In the present case of chemotactic dynamics with logistic growth, namely Eq.~\eqref{eq:Langevin}, even if an exact exponent identity holds as a consequence of the pseudo-Galilean transformation~\eqref{eq:Galilean} (see Eq.~\eqref{eq:expidentity}), the noise term $\mathcal{D}_0$ acquires a non-trivial renormalization due to the presence of $\lambda$. As a consequence, the critical exponents are no longer known exactly, but we describe in the following how they can be calculated by using RG techniques.

\section{Renormalization group analysis}\label{Sec:RG}

\subsection{Perturbative expansion}

Transforming Eq.~\eqref{eq:Langevin} in Fourier space\footnote{Here we use the following convention 
$f(\mathbf{x},t) = \int_{\hat{k}} f(\mathbf{k}, \omega) \ee^{-\ii \omega t + \ii \mathbf{k} \cdot \mathbf{x} }$ with shorthand notation
$\hat{k} \equiv (\mathbf{k}, \omega)$ and
$\int_{\hat{k}} \equiv \int {\rm{d}}^d \mathbf{k} \, {\rm{d}} \omega / (2\pi)^{d+1}$.}
 we obtain
\begin{equation}\label{eq:LangevinFourier}
    \rho (\hat{k}) = G_0 (\hat{k}) \zeta  (\hat{k}) + G_0 (\hat{k}) \int_{\hat{k}_1} V_0(\mathbf{k}, \mathbf{k}_1) \rho (\hat{k}_1)\rho (\hat{k}-\hat{k}_1) \,,
\end{equation}
where $G_0 (\hat{k}) \equiv (- \ii \tau \omega + D k^2 + \sigma)^{-1}$
is the bare propagator, while the nonlinearities $\mu_{1,2}$ and $\lambda$ enter via the the vertex function $V_0 (\mathbf{k}, \mathbf{k}_1)$, which upon symmetrization $\mathbf{k}_1 \leftrightarrow \mathbf{k} - \mathbf{k}_1$, reads
\begin{equation} \label{eq:bareVertex}
\begin{split}
    V_0 (\mathbf{k}, \mathbf{k}_1) = & \, -\lambda - \mu_2 \, \frac{ \mathbf{k}_1\cdot(\mathbf{k}-\mathbf{k}_1)k^2}{(\mathbf{k}-\mathbf{k}_1)^2k_1^2} + \\ 
    & + \frac{1}{2}\mu_1\left[\frac{\mathbf{k}\cdot\mathbf{k}_1}{k_1^2}
    + \frac{\mathbf{k}\cdot(\mathbf{k}-\mathbf{k}_1)}{(\mathbf{k}-\mathbf{k}_1)^2} \right] \,.
    \end{split}
\end{equation}
We then denote the bare correlation function (spectral function) $N_0(\hat{k}) = 2 \mathcal{D}_0 |G_0(\hat{k})|^2$. Note that we have discarded the conserving part of the noise correlations~\eqref{eq:zeta-corr-full} proportional to $\mathcal{D}_2$, as it is irrelevant in the RG sense. 
In view of setting the perturbative expansion, it is convenient to introduce the following diagrammatic representation~\cite{forster1977largedistance,medina1989burgers}
\begin{equation}\label{eqs:barequantities}
G_0 (\hat{k}) =
\begin{tikzpicture}[baseline=-.1cm]
    \draw [-To] (0,0)   -- (1/4,0);
    \draw       (1/4,0) -- (1/2,0);
\end{tikzpicture}
~,~~~~ V_0 (\mathbf{k},\mathbf{k}_1) =
\begin{tikzpicture}[baseline=-.1cm]
    \draw       (0,0)   -- (1/4,0);
    \draw       (1/4,0) -- (1/2,1/4);
    \draw       (1/4,0) -- (1/2,-1/4);
\end{tikzpicture}
~,~~~~ N_0 (\hat{k}) =
\begin{tikzpicture}[baseline=-.1cm]
    \draw [-To] (0,0)         -- (1/4,0);
    \draw       (1/4,0)       -- (1/4+1/5,0);
    \draw [To-] (1/4+1/5,0)   -- (1/2+1/5,0);
    \filldraw [gray!50] (1/4+1/10,0) circle (2pt);
    \draw (1/4+1/10,0) circle (2pt);
\end{tikzpicture}~.
\end{equation}
The integral equation \eqref{eq:LangevinFourier} is then solved perturbatively in the vertex $V_0$: the one-loop corrections to the bare quantities \eqref{eqs:barequantities} are schematically given by
\begin{align}
G (\hat{k}) & = ~~
\begin{tikzpicture}[baseline=-.1cm]
    \draw [-To] (0,0)   -- (1/4,0);
    \draw       (1/4,0) -- (1/2,0);
\end{tikzpicture}
\, + \, 4 ~
\begin{tikzpicture}[baseline=-.1cm]
    \draw [-To]  (0,0)   -- (3/16,0);
    \draw        (3/16,0)   -- (1/4,0);
    \draw [-To]  (1,0) -- (1+3/16,0);
    \draw        (1+3/16,0) -- (5/4,0);
    \draw [->]   (1/4,0) arc (180:270:3/8);
    \draw        (5/8,-3/8) arc (270:360:3/8);
    \draw [->]   (1/4,0) arc (180:100:3/8);
    \draw [->]   (1,0) arc (0:80:3/8);
    \filldraw [gray!50] (5/8,3/8) circle (2pt);
    \draw (5/8,3/8) circle (2pt);
\end{tikzpicture}~, \nonumber
\\
V(\mathbf{k},\mathbf{k}_1) & = ~~
\begin{tikzpicture}[baseline=-.1cm]
    \draw       (0,0)   -- (1/4,0);
    \draw       (1/4,0) -- (1/2,1/4);
    \draw       (1/4,0) -- (1/2,-1/4);
\end{tikzpicture}
\, + \, 4 ~
\begin{tikzpicture}[baseline=-.1cm]
    \draw        (0,0)   -- (1/4,0);
    \draw        (0.89,-0.26) -- (1.14,-0.51);
    \draw        (0.89,0.26) -- (1.14,0.51);
    \draw [->]   (1/4,0)    arc (180:215:3/8);
    \draw [->]   (5/8,3/8)  arc (90:0:3/8);
    \draw [->]   (1/4,0)    arc (180:90:3/8);
    \draw [->]   (1,0) arc (360:235:3/8);
    \filldraw [gray!50] (0.35,-0.26) circle (2pt);
    \draw (0.35,-0.26) circle (2pt);
\end{tikzpicture}
\, + \, 4 ~
\begin{tikzpicture}[baseline=-.1cm]
    \draw        (0,0)   -- (1/4,0);
    \draw        (0.89,-0.26) -- (1.14,-0.51);
    \draw        (0.89,0.26) -- (1.14,0.51);
    \draw [->]   (1/4,0)    arc (180:270:3/8);
    \draw [->]   (1,0)  arc (0:125:3/8);
    \draw [->]   (1/4,0)    arc (180:150:3/8);
    \draw [->]   (5/8,-3/8) arc (270:360:3/8);
    \filldraw [gray!50] (0.35,0.26) circle (2pt);
    \draw (0.35,0.26) circle (2pt);
\end{tikzpicture}
\, + \, 4 ~
\begin{tikzpicture}[baseline=-.1cm]
    \draw        (0,0)   -- (1/4,0);
    \draw        (0.89,-0.26) -- (1.14,-0.51);
    \draw        (0.89,0.26) -- (1.14,0.51);
    \draw [->]   (1/4,0)    arc (180:270:3/8);
    \draw [->]   (5/8,3/8)  arc (90:10:3/8);
    \draw [->]   (1/4,0)    arc (180:90:3/8);
    \draw [->]   (5/8,-3/8) arc (270:350:3/8);
    \filldraw [gray!50] (1,0) circle (2pt);
    \draw (1,0) circle (2pt);
\end{tikzpicture}\nonumber \,,
\\
N (\hat{k}) & = \,
\begin{tikzpicture}[baseline=-.1cm]
    \draw [-To] (0,0)         -- (1/4,0);
    \draw       (1/4,0)       -- (1/4+1/5,0);
    \draw [To-] (1/4+1/5,0)   -- (1/2+1/5,0);
    \filldraw [gray!50] (1/4+1/10,0) circle (2pt);
    \draw (1/4+1/10,0) circle (2pt);
\end{tikzpicture}
\, + \, 2 ~
\begin{tikzpicture}[baseline=-.1cm]
    \draw [-To]  (0,0)   -- (3/16,0);
    \draw        (3/16,0)   -- (1/4,0);
    \draw        (1,0) -- (1+1/16,0);
    \draw [To-]  (1+1/16,0) -- (5/4,0);
    \draw [->]   (1/4,0) arc (180:100:3/8);
    \draw [->]   (1,0) arc (0:80:3/8);
    \draw [->]   (1,0) arc (0:-80:3/8);
    \draw [->]   (1/4,0) arc (180:260:3/8);
    \filldraw [gray!50] (5/8,3/8) circle (2pt);
    \draw (5/8,3/8) circle (2pt);
    \filldraw [gray!50] (5/8,-3/8) circle (2pt);
    \draw (5/8,-3/8) circle (2pt);
\end{tikzpicture}~, \nonumber 
\end{align}
where the prefactors in front of each one-loop contribution accounts for all the possible noise contractions (the corresponding explicit expressions are given in Supplementary Material).
The renormalization group procedure can then be set by first integrating out fluctuations within the momentum shell $\Lambda \ee^{-\ell} \leq |\mathbf{k}| \leq \Lambda$, where $\Lambda$ represents the momentum cutoff (set, e.g.\,, by the inverse particle size).
This momentum integration is followed by a rescaling back to the original cutoff $\Lambda$, a rescaling of the fields according to \mbox{$\rho \to \rho' = \ee ^ {\ell \chi} \rho$}, \mbox{$\eta \to \eta' = \ee ^ {-\ell \tilde\chi} \eta$}, and a temporal rescaling  \mbox{$t \to t' = \ee^{\ell z}t$}.
The roughness exponent $\chi$, the dynamical exponent $z$, and the exponent $\tilde\chi$, characterize the scaling behavior of the critical system in the macroscopic limit. For instance, the long-time and large-scale particle density correlations assume the scaling form~\cite{tauber2014critical,medina1989burgers}
\begin{equation}
  \langle \rho(\mathbf{x},t)\rho(\mathbf{x}',t) \rangle \sim |\mathbf{x}'-\mathbf{x}|^{2\chi} F\left(\frac{|t'-t|}{|\mathbf{x}'-\mathbf{x}|^z}\right) \, ,
\end{equation}
where $F$ is a scaling function.

The exponent $\tilde\chi$ determines the scaling properties of the response function $R$ of the system to an external force $h(\mathbf{x}',t')$. Namely, by defining \mbox{$R(\mathbf{x},t,\mathbf{x}',t') = \delta \langle \rho(\mathbf{x},t) \rangle/\delta h(\mathbf{x}',t') |_{h=0}$}, we have
\mbox{$R(\mathbf{x},t,\mathbf{x}',t') \sim |\mathbf{x}'-\mathbf{x}|^{\chi+\tilde\chi} H(|t'-t|/|\mathbf{x}'-\mathbf{x}|^z)$} in the asymptotic limit, with $H$ a scaling function.
The three critical exponents $z$, $\chi$ and $\tilde\chi$ are in principle independent; however, the pseudo-Galilean symmetry~\eqref{eq:Galilean} enforces the exact exponent identity~\cite{mahdisoltani2021nonequilibrium}
\begin{align}\label{eq:expidentity}
  z+\chi = 0 \, .
\end{align}
Futhermore, we note that for systems relaxing to an  equilibrium configuration (which is not the case here), the fluctuation-dissipation theorem provides a relation between response function and correlation function~\cite{aron2010symmetries}, which implies the exponent identity 
$2\chi -z = \chi+\tilde\chi$. 
This identity is broken\footnote{Note that the latter identity is accidentally recovered at the fixed points of the dynamics when the nonlinear growth coupling is set to zero ($\lambda = 0$) -- even though the polarity-induced coupling is itself a  nonequilibrium interaction and is not derivable from a free-energy functional~\cite{mahdisoltani2021nonequilibrium}.} for the critical dynamics of Eq.~\eqref{eq:Langevin}. 

We obtain the RG functions in the limit $\ell \to 0$ and through a dimensional expansion in the parameter $\varepsilon = d_c - d \ll 1$ around the upper critical dimension $d_c = 6$. 
On defining the rescaled variables $\bar\lambda^2 = \lambda^2 K_d\Lambda^{-\varepsilon} \ddd_0 / (D^3\tau)$ and $\bar\mu^2_{1,2} = \mu^2_{1,2} K_d\Lambda^{-\varepsilon} \ddd_0 / (D^3\tau)$,
where $K_d \equiv S_d/(2\pi)^d$ with $S_d$ the area of a unit sphere in $d$ dimensions, we obtain the following RG flow equations
\begingroup%
\allowdisplaybreaks%
\begin{subequations}\label{eq:RGfunctions}%
\begin{align}%
 \frac{\partial_\ell \mathcal{D}_0}{\mathcal{D}_0} =\, &
 d+z+2\tilde\chi + \bar\lambda^2 \,, \\
\begin{split}
 \frac{\partial_\ell D}{D} =\, &
 d+z+\chi+\tilde\chi-2
 - \frac{25}{48}  \bar\mu_1^2
 - \frac{15}{8}   \bar\mu_1\bar\mu_2
\\
 &
 - \frac{1}{16}   \bar\mu_1\bar\lambda  
 + \frac{53}{24}  \bar\mu_2\bar \lambda
+ \frac{2}{3} \bar\lambda^2 \, , 
\end{split}\\
 \partial_\ell \tau =\, &
 (d+\chi+\tilde\chi)\tau - \frac{5}{6}\bar\mu_1\bar\lambda + \bar\lambda^2 \, , \\
\begin{split}
 \partial_\ell {\bar\mu}_1 =\, &
 \left( d+z+2\chi+\tilde\chi \right){\bar\mu}_1 -
 \frac{7}{8}    \bar\mu_1^2 \bar\lambda
 + \frac{11}{6} \bar\mu_1\bar\mu_2 \bar\lambda  \\
 &
 + \frac{1}{2}  \bar\mu_2^2 \bar\lambda
 + \bar\mu_1\bar\lambda ^2
 - 2\bar\mu_2\bar\lambda^2 \, ,
\end{split}\\
\begin{split}
 \partial_\ell \bar\mu_2 =\, &
 \left( d+z+2\chi+\tilde\chi \right){\bar\mu}_2 -
 \frac{1}{48}   \bar\mu_1^2 \bar\lambda
  \\
 &
 + \frac{1}{12} \bar\mu_1\bar\mu_2 \bar\lambda + \frac{1}{4}  \bar\mu_2^2 \bar\lambda \, ,
\end{split}\\
\begin{split}
 \partial_\ell {\bar\lambda} =\, &
 (d+z+2\chi+\tilde\chi) \bar\lambda
 + \frac{35}{16}\bar\mu_1^2 \bar\lambda
 + \frac{5}{4}\bar\mu_1\bar\mu_2 \bar\lambda  \\
 &
 + \frac{5}{12}\bar\mu_2^2 \bar\lambda
 - 6 \bar\mu_1\bar\lambda^2
 - \frac{4}{3}\bar\mu_2\bar\lambda^2
 + 4 \bar\lambda^3 \, .
\end{split}
 \end{align}%
 \end{subequations}%
\endgroup%
By taking the logarithm of the definition of $\bar\mu_{1,2}$ and $\bar\lambda$, differentiating w.r.t. $\ell$, and taking advantage of Eq.~\eqref{eq:RGfunctions}, we obtain the following closed flow equations for arbitrary $\chi$, $\tilde\chi$ and $z$,
\begin{subequations}\label{eq:reducedfloweqs}%
\begin{align}%
\begin{split}
\partial_\ell \bar\lambda =\, & \frac{1}{2} \bar\lambda  \varepsilon
+ 3               \bar\lambda ^3
- \frac{527}{96}  \bar\mu _1 \bar\lambda ^2
- \frac{223}{48}  \bar\mu _2 \bar\lambda ^2 \\
&
+ \frac{95}{32}   \bar\mu _1^2 \bar\lambda + \frac{5}{12}    \bar\mu _2^2 \bar\lambda
+ \frac{65}{16}   \bar\mu _1 \bar\mu _2 \lambda \,,
\end{split}\\
\begin{split}
\partial_\ell \bar\mu_1 =\, & \frac{1}{2}\bar\mu _1 \varepsilon
+ \frac{25}{32}   \bar\mu _1^3
- \frac{35}{96}   \bar\mu _1^2 \bar\lambda
+ \frac{45}{16}   \bar\mu _1^2 \bar\mu _2 \\
&
- \frac{71}{48}   \bar\mu _2 \bar\mu _1 \bar\lambda 
+ \frac{1}{2}     \bar\mu_2^2 \bar\lambda
- 2               \bar\mu _2 \bar\lambda ^2 \,,
\end{split}\\
\begin{split}
\partial_\ell \bar\mu_2 =\, & \frac{1}{2} \bar\mu _2 \varepsilon
- \frac{49}{16}     \bar\mu _2^2 \bar\lambda
+ \frac{45}{16}     \bar\mu _1 \bar\mu _2^2
-                   \bar\lambda ^2 \bar\mu _2  \\
&
+ \frac{19}{32}     \bar\mu_1 \bar\mu _2 \bar\lambda
+ \frac{25}{32}     \bar\mu _1^2 \bar\mu _2
- \frac{1}{48}      \bar\mu _1^2 \bar\lambda
\,.
\end{split}
\end{align}%
\end{subequations}%
These one-loop RG equations are the main technical result of this paper.
Consistently with the proximity to the Gaussian fixed point, we choose the values of $\chi$, $\tilde\chi$ and $z$ such that $D$, $\mathcal{D}_0$ and $\tau$ do not flow under RG, i.e.\,, we impose $\partial_\ell D = \partial_\ell \mathcal{D}_0 = \partial_\ell \tau = 0$. This choice is such that the scaling of the correlation length $\xi$, which is controlled by the ratio $D/\sigma$, is determined solely by the scaling of $\sigma$.
Likewise, the density fluctuations in the renormalized system, which are controlled by the ratio $\mathcal{D}_0/(D \tau)$, remain unchanged at the critical point and their scaling can therefore be determined directly from $\chi$, $\tilde\chi$ and $z$.
Adopting this choice, we find
\begin{subequations}%
\begin{align}%
z & = 2
\!+\! \frac{25}{48}   \bar\mu _1^2
\!+\! \frac{15}{8}    \bar\mu _1 \bar\mu _2
\!-\! \frac{37}{48}    \bar\mu_1 \bar\lambda
\!-\! \frac{53}{24}   \bar\mu_2 \bar\lambda
\!+\! \frac{1}{3}     \bar\lambda ^2 \,,
\\
\tilde\chi & = -\frac{1}{2} (z +d + \bar\lambda ^2) \, , \\
\chi & = -d -\tilde\chi -\bar\lambda ^2 + \frac{5}{6} \bar\mu_1 \bar\lambda \, .
\end{align} \label{eq_exponents}%
\end{subequations}%

Lastly, the critical state of the system can be reached by fine-tuning $\sigma$ to its critical value $\sigma_c$. This is a fluctuation-induced and non-universal quantity, the value of which we report in the Supplementary Material.
It is customary to introduce the reduced control parameter $r \equiv \sigma - \sigma_c$ such that the critical state of the system is reached for $r \to 0$. The corresponding  RG flow then reads 
\begin{equation}
\frac{\partial_\ell r}{ r } = \label{eq:mass}
d+z+\chi+\tilde\chi - \frac{7}{2} \bar\mu_1\bar\lambda  + \frac{1}{3}\bar\mu_2 \bar\lambda + 4 \bar\lambda^2 \,.
\end{equation}
%

\subsection{Fixed points}

\begin{figure}[h!]
\centering
	\begin{minipage}[c]{.8\linewidth}
		\centering
		\includegraphics[width=\linewidth]{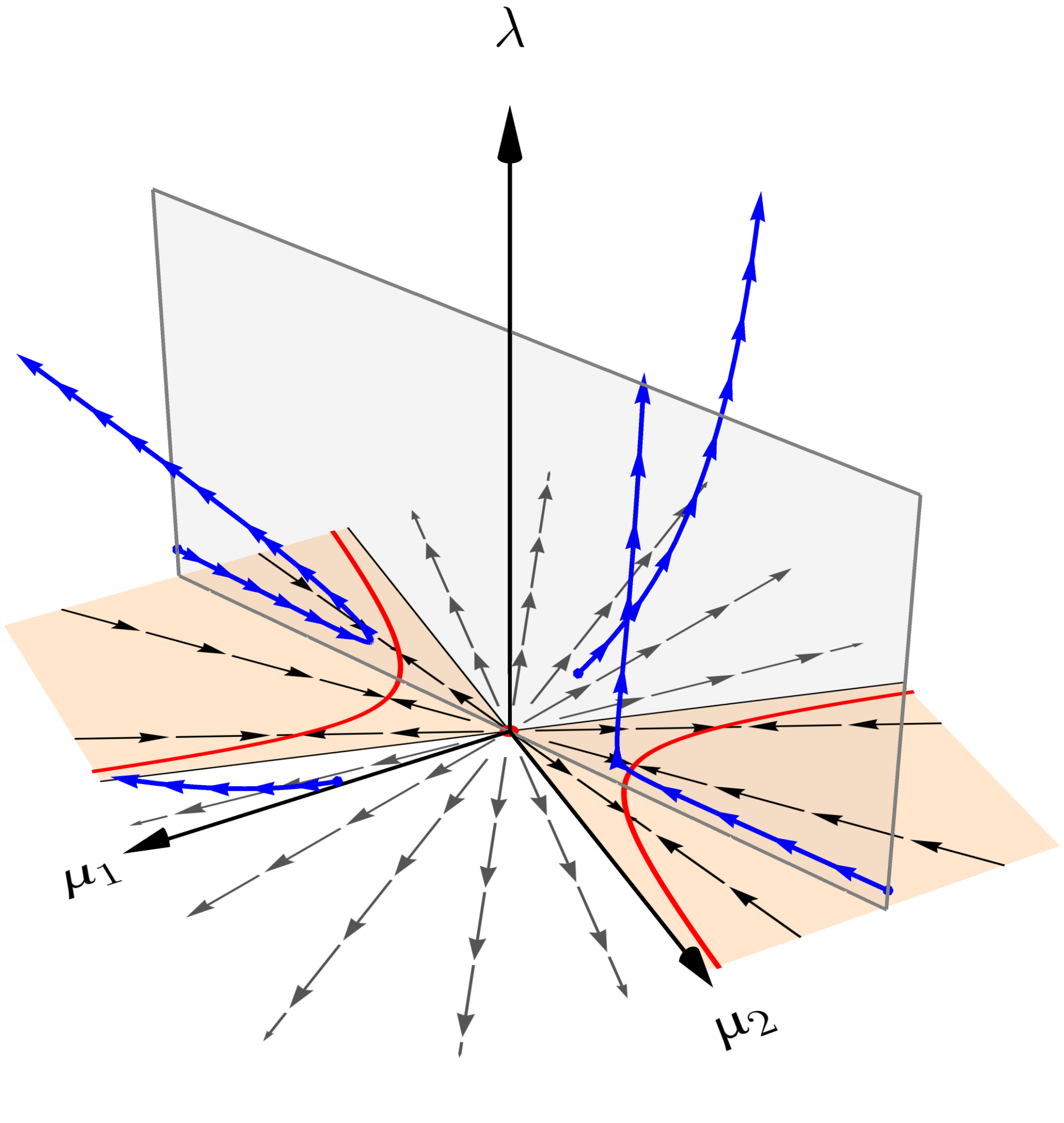}
	\end{minipage}  
	\\
	\begin{minipage}[c]{.8\linewidth}
		\centering
		\includegraphics[width=\linewidth]{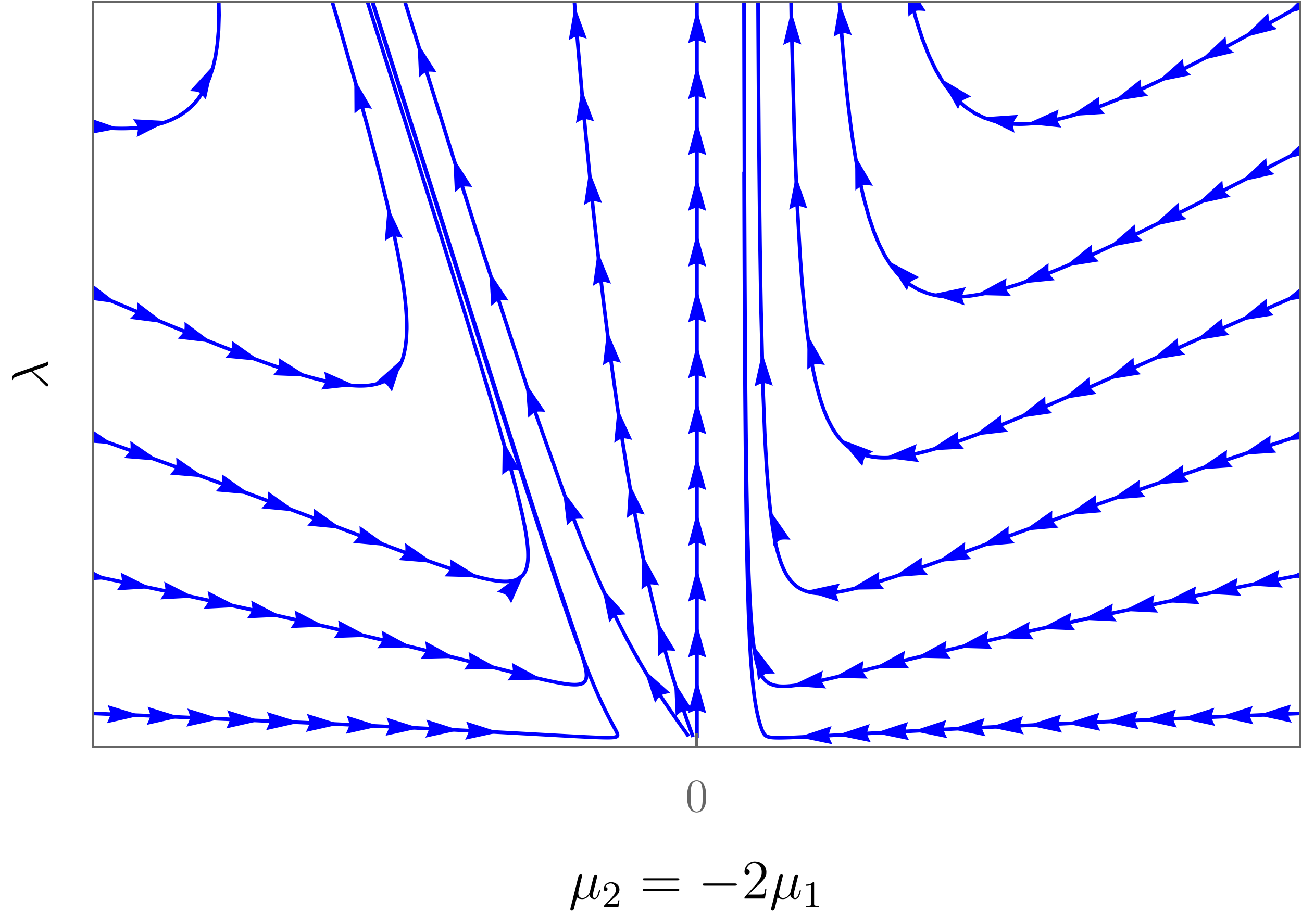}
	\end{minipage} 
	\caption{%
	\textbf{(Top)} One-loop RG flows in the parameter space. The red hyperbolas in the $\lambda=0$ plane correspond to the fixed-point solutions in Eq.~\eqref{eq:FPconserved}. The blue arrows show four instances of RG flow lines with $\lambda \neq 0$, all appearing to exhibit runaway behavior.  
	\textbf{(Bottom)} Projected RG flow on the plane $\mu_2=-2\mu_1$ (gray plane in the top panel).
    Note the crossover behavior of the trajectories starting from $\lambda \neq 0$, which first approach the line of fixed points at $\lambda=0$ before diverging.
	}
\label{Fig:conservedflow}
\end{figure}

The system of Eqs.~\eqref{eq:reducedfloweqs} admits the Gaussian fixed point \mbox{$\bar{\mu}_1 ^\star = \bar{\mu}_2 ^\star = \bar{\lambda} ^\star = 0$} as a solution, which is unstable below the upper critical dimension $d<d_c=6$ and stable above it.
In the parameter space, we then focus on the plane  \mbox{$(\bar{\mu}_1,\bar{\mu}_2,\bar{\lambda}^\star = 0)$}, for which Eqs.~\eqref{eq:reducedfloweqs} reduce to a closed system in the variables $(\bar{\mu}_1,\bar{\mu}_2)$.
The RG flow on this plane is governed by the stable lines of fixed points defined by the conics
\begin{equation}\label{eq:FPconserved}
  \bar\mu _2 ^\star = - \frac{8 \varepsilon}{45 \bar\mu _1 ^\star} - \frac{5 \bar\mu _1 ^\star}{18} \hspace{10pt} \text{and} \hspace{10pt} \bar\lambda ^\star = 0 \,.
\end{equation}
The asymptotic lines corresponding to these hyperbolas separate regions of the reduced couplings plane where the RG flow runs away to infinity, shown by the blue arrows in the top panel of Fig.~\ref{Fig:conservedflow}, from the basin of attraction of the fixed points that are shaded in red in the same figure.   
(See Ref.~\cite{mahdisoltani2021nonequilibrium} for a detailed analysis of the flow equations of the chemotactic model without the nonlinear growth term.) 

As discussed in Ref.~\cite{mahdisoltani2021nonequilibrium}, the  runaway behavior of the one-loop RG flows in the $\lambda = 0$ plane could hint to several different scenarios, and thus it calls for further investigations.   
In particular, it may be an artifact of the one-loop approximation used here. On the other hand, the runaway solution could signal a first-order phase transition, or, alternatively, it may suggest that the critical behavior of the dynamics is ruled by a strong-coupling fixed point that is not within the reach of the perturbative RG calculations. 
The latter possibility is supported by a connection between the chemotactic Langevin dynamics~\eqref{eq:Langevin} and the KPZ equation~\cite{kardar1986dynamic}, which describes the fluctuations of a growing interface and is known to exhibit a nonperturbative behavior~\cite{wiese1998perturbation,kloss2014kardarparisizhang}. We further discuss this connection below.

For chemotactic dynamics with logistic growth, our one-loop analysis shows that no stable real-valued RG fixed point exists in the space $(\bar{\mu}_1,\bar{\mu}_2,\bar{\lambda}\neq0)$ and in the vicinity of the upper critical dimension (see the Supplementary Material for a discussion of possible fixed points at lower dimensions).
A linear stability analysis which considers a perturbation around the fixed-point solutions~\eqref{eq:FPconserved} gives the eigenvalues $\uptheta_{1,2,3}$ of the associated stability matrix as
\begin{align}
    \uptheta_1 = - \varepsilon \, , \quad \uptheta_2 = 0 \, , \quad \uptheta_3 = \frac{16-220\bar\mu_1^2 + 2275\bar\mu_1^4}{1215 \bar\mu_1^2} \varepsilon  \, .
\end{align}
The eigenvalue $\uptheta_1$ is negative for $\varepsilon>0$ and indicates that the line of fixed points is attractive in the plane $\lambda=0$, as already discussed above.
The eigenvalue $\uptheta_2=0$ is the marginal direction along the line of fixed points. 
The eigenvalue $\uptheta_3$ is always positive for $\varepsilon>0$ and with $\bar\mu _1\in \mathbb{R}$, and it indicates that the perturbations in a direction orthogonal to the plane $\lambda=0$ are unstable.

The analysis presented above therefore shows the absence of stable perturbative fixed point for the chemotactic system with growth.
This contrasts with the findings of Ref.~\cite{gelimson2015collective} where a stable fixed point was reported for the dynamics of dividing chemotactic cells without the polarity-induced coupling.
Importantly, even if not present at the microscopic level, the polarity-induced chemotactic term proportional to $\mu_2$ is immediately generated by the presence of the growth term $\propto\lambda$ (see Eqs.~\eqref{eq:reducedfloweqs}).
This observation highlights the fundamental role played by the  polarity-induced mechanism in describing the macroscopic properties of dividing chemotactic colonies.

In passing, we mention that along the line $(\bar{\mu}_1^\star  = 0, \bar{\mu}_2^\star  = 0, \bar{\lambda})$, we find two imaginary RG fixed points with coordinates $\bar\lambda ^\star = \pm \ii \sqrt{\varepsilon/6}$.
These correctly describe the dynamics of the Yang--Lee edge singularity~\cite{yang1952statistical,lee1952statistical} for which, at one-loop order, we recover the known critical exponents $z = 2- \varepsilon/18 $ and $\chi = - 2 + 5\varepsilon/9 $~\cite{fisher1978yanglee,breuer1981equation}.

As a final remark, we discuss the relation between Eq.~\eqref{eq:Langevin} and the KPZ dynamics~\cite{kardar1986dynamic}.
Consider the roughening of an interface with height profile $h(\mathbf{r},t)$ and subject to a long-ranged correlated noise $\tilde\eta$.
The corresponding dynamics reads~\cite{medina1989burgers}
\begin{align}\label{eq:kpz}
    \partial_t h = D \nabla^2 h + g (\nabla h)^2 + \tilde\eta \, ,
\end{align}
where the noise $\tilde\eta$ has correlations $\langle \tilde\eta(\mathbf{k},\omega) \tilde\eta(\mathbf{k}_1,\omega_1) \rangle = 2 \mathcal{D}_{\rm LR} k^{-2\theta} \delta^d(\mathbf{k} + \mathbf{k}_1)\delta(\omega+\omega_1)$,
with $\theta$ characterizing the long-range nature of its spatial correlations (the original short-ranged KPZ equation is recovered for $\theta=0$).
By taking the Laplacian of Eq.~\eqref{eq:kpz}, one then arrives at 
\begin{align}\label{eq:laplacian_kpz}
    \partial_t \nabla^2 h = D \nabla^2 ( \nabla^2 h )+ g \nabla^2(\nabla h)^2 + \eta \, ,
\end{align}
where the noise $\eta$ is now characterized by
%
    $\langle \eta(\mathbf{k},\omega) \eta(\mathbf{k}_1,\omega_1) \rangle = 2 \mathcal{D}_{\rm LR} k^{4-2\theta} \delta^d(\mathbf{k} + \mathbf{k}_1)\delta(\omega + \omega_1) \, .$
%
In the absence of the nonlinearities $\mu_1$ and $\lambda$, the Langevin equation~\eqref{eq:Langevin}, which now describes the dynamics of chemotactic particles with only the polarity-induced mechanism, can be mapped to the KPZ model~\eqref{eq:laplacian_kpz} with long-range correlated noise, by setting $\theta=2$ and   identifying $\phi \leftrightarrow h$ and $\mu_2 \leftrightarrow g$.
The pseudo-Galilean symmetry~\eqref{eq:Galilean} of the original chemotactic model then corresponds to the Galilean symmetry of the KPZ equation. 
Note that in this case, Eq.~\eqref{eq_exponents} yields the critical exponents $z=2$, $\chi=-2-\varepsilon/2$ and $\tilde\chi=-4+\varepsilon/2$,
which are consistent with the fact that these exponents should not acquire correction at any order of the perturbation theory with $\varepsilon$-expansion (rather than a ``fixed-$d$'' scheme)~\cite{wiese1998perturbation,lassig1995renormalization}.  

\section{Conclusion}

Starting from a microscopic model of chemotactic particles undergoing birth and death processes, we have derived a coarse-grained, field-theoretical  description for the dynamics of density fluctuations in a growing self-chemotactic colony (see Eq.~\eqref{eq:Langevin}).
In this Langevin description, we identify a pseudo-Galilean symmetry, which is also present in the case of non-dividing chemotactic particles~\cite{mahdisoltani2021nonequilibrium}; 
this symmetry is reminiscent of the Galilean symmetry of the Kardar--Parisi--Zhang equation~\cite{kardar1986dynamic}. 
We then use perturbative RG techniques to investigate the scaling behavior of the dynamics at its critical point. 
Crucially, our analysis reveals that the interplay of the fluctuations effects in the presence of the Keller--Segel chemotactic interactions and the logistic growth nonlinearity, generate the so-called polarity-induced chemotactic coupling ($\propto \mu_2$ in Eq.~\eqref{eq:Langevin}) at macroscopic scales. Although the microscopic origin of this unconventional coupling may be attributed to particle polarity effects in chemotactic dynamics~\cite{mahdisoltani2021nonequilibrium}, its generation along the RG flow in the presence of growth processes shows that it needs to be incorporated in models of dividing chemotactic particles, even if the cells have an isotropic gradient-sensing mechanism with no polarity effects at the microscopic level.

We show that including the polarity-induced coupling changes the critical picture of the system compared to the analysis of Ref.~\cite{gelimson2015collective}. In particular, the chemotactic lines of fixed points that exist in the absence of the growth coupling $\lambda$~\cite{mahdisoltani2021nonequilibrium} are unstable in the direction of $\lambda$, and we report the absence of any stable fixed point of the one-loop perturbative RG flow near the upper critical dimension $d_c=6$ (see the Supplemental Material for the discussion of a stable fixed point in $d=2$).
Importantly, even in the absence of a stable fixed point, the behavior of large but finite systems can still be dictated by the presence of the line of fixed points in the plane $\lambda=0$, as illustrated by the crossover behavior in Fig.~\ref{Fig:conservedflow}.

Determining whether a stable perturbative fixed point would be found at higher order in the perturbative expansion, or whether the absence of a stable perturbative fixed point is the signature of a strong-coupling critical point that would govern the critical behavior, requires further investigation,  e.g., using nonperturbative RG techniques~\cite{canet2011nonperturbative,duclut2017frequency,duclut2017nonuniversality,nprg-rev}.

\acknowledgments
R.B.A.Z. and C.D. thank Davide Squizzato for illuminating discussions. 
RBAZ acknowledges the support from the French ANR through the project NeqFluids (grant ANR-18-CE92-0019).
S.M. acknowledges support from the University of Oxford through the Clarendon
Fund and St John's College Kendrew Scholarship, and the Rudolf Peierls Centre for Theoretical Physics.
A.G. acknowledges support from the MIUR PRIN Project “Coarse-grained description for nonequilibrium systems and transport phenomena (CO-NEST),” No. 201798CZL. 
R.G. acknowledges support from the Max Planck School Matter to Life and the MaxSynBio Consortium which are jointly funded by the Federal Ministry of Education and Research (BMBF) of Germany and the Max Planck
Society.

\appendix

\clearpage

\onecolumngrid
\section{Details of the renormalization group analysis}\label{Sec:Appendix}

In this section, following standard renormalization group (RG) methods~\cite{tauber2014critical}, we compute the one-loop corrections to the bare quantities in Eq.~\eqref{eqs:barequantities}, which were outlined in the main text. Adopting the following notation $\int_{\hat q}^>\equiv\int_{-\infty}^\infty\dd \omega \int_{\Lambda/b\leq |\bm q| \leq \Lambda}\dd^d \bm q$, where $\Lambda$ and $b$ were introduced in the main text, we have
\begin{subequations}%
\begin{align}
G_1(\hat k) & =
\begin{tikzpicture}[baseline=-.1cm]
    \draw        (0,0)   -- (1/4,0);
    \draw        (1,0) -- (5/4,0);
    \draw [->]   (1/4,0) arc (180:270:3/8);
    \draw        (5/8,-3/8) arc (270:360:3/8);
    \draw [->]   (1/4,0) arc (180:100:3/8);
    \draw [->]   (1,0) arc (0:80:3/8);
    \filldraw [gray!50] (5/8,3/8) circle (2pt);
    \draw (5/8,3/8) circle (2pt);
\end{tikzpicture} \nonumber\\
    & = \int^>_{\hat{q}}  N_0 (\hat{k}/2 +\hat{q})
                          G_0 (\hat{k}/2 - \hat{q})
                          V_0 (\mathbf{k}, \mathbf{k}/2+ \mathbf{q})
                          V_0 (\mathbf{k}, \mathbf{k}/2 - \mathbf{q}) \, , \\
V_1^{\rm (a)}(\hat k, \hat k/2 + \hat k_1) & = \begin{tikzpicture}[baseline=-.1cm]
    \draw        (0,0)   -- (1/4,0);
    \draw        (0.89,-0.26) -- (1.14,-0.51);
    \draw        (0.89,0.26) -- (1.14,0.51);
    \draw [->]   (1/4,0)    arc (180:215:3/8);
    \draw [->]   (5/8,3/8)  arc (90:0:3/8);
    \draw [->]   (1/4,0)    arc (180:90:3/8);
    \draw [->]   (1,0) arc (360:235:3/8);
    \filldraw [gray!50] (0.35,-0.26) circle (2pt);
    \draw (0.35,-0.26) circle (2pt);
\end{tikzpicture} \nonumber\\
    & =  \int^>_{\hat{q}}  N_0 (\hat{k}/2 - \hat{q})
                          G_0 (\hat{k}/2 + \hat{q})
                          G_0 (\hat{q} - \hat{k}_1)
                          V_0 (\mathbf{k}, \mathbf{k}/2 + \mathbf{q})
                          V_0 (\mathbf{k}/2 + \mathbf{q}, \mathbf{k}/2 + \mathbf{k}_1)
                          V_0 (\mathbf{q} - \mathbf{k}_1, \mathbf{k}/2 - \mathbf{k}_1) \, ,\\
V_1^{\rm (b)}(\hat k, \hat k/2 + \hat k_1)  & = \begin{tikzpicture}[baseline=-.1cm]
    \draw        (0,0)   -- (1/4,0);
    \draw        (0.89,-0.26) -- (1.14,-0.51);
    \draw        (0.89,0.26) -- (1.14,0.51);
    \draw [->]   (1/4,0)    arc (180:270:3/8);
    \draw [->]   (1,0)  arc (0:125:3/8);
    \draw [->]   (1/4,0)    arc (180:150:3/8);
    \draw [->]   (5/8,-3/8) arc (270:360:3/8);
    \filldraw [gray!50] (0.35,0.26) circle (2pt);
    \draw (0.35,0.26) circle (2pt);
\end{tikzpicture} \nonumber\\
    & =  \int^>_{\hat{q}}  N_0 (\hat{k}/2 + \hat{q})
                          G_0 (\hat{k}/2 - \hat{q})
                          G_0 (\hat{k}_1 - \hat{q})
                          V_0 (\mathbf{k}, \mathbf{k}/2 + \mathbf{q})
                          V_0 (\mathbf{k}/2 - \mathbf{q}, \mathbf{k}_1 - \mathbf{q})
                          V_0 (\mathbf{k}_1 - \mathbf{q}, \mathbf{k}/2 + \mathbf{k}_1) \, ,\\
V_1^{\rm (c)}(\hat k, \hat k/2 + \hat k_1) & = \begin{tikzpicture}[baseline=-.1cm]
    \draw        (0,0)   -- (1/4,0);
    \draw        (0.89,-0.26) -- (1.14,-0.51);
    \draw        (0.89,0.26) -- (1.14,0.51);
    \draw [->]   (1/4,0)    arc (180:270:3/8);
    \draw [->]   (5/8,3/8)  arc (90:10:3/8);
    \draw [->]   (1/4,0)    arc (180:90:3/8);
    \draw [->]   (5/8,-3/8) arc (270:350:3/8);
    \filldraw [gray!50] (1,0) circle (2pt);
    \draw (1,0) circle (2pt);
\end{tikzpicture} \nonumber\\
    & =  \int^>_{\hat{q}}  N_0 (\hat{k}_1 - \hat{q})
                          G_0 (\hat{k}/2 + \hat{q})
                          G_0 (\hat{k}/2 - \hat{q})
                          V_0 (\mathbf{k}, \mathbf{k}/2 + \mathbf{q})
                          V_0 (\mathbf{k}/2 + \mathbf{q}, \mathbf{k}/2 + \mathbf{k}_1)
                          V_0 (\mathbf{k}/2 - \mathbf{q}, \mathbf{k}/2 - \mathbf{k}_1) \,, \\
N_1(\hat{k}) & = \begin{tikzpicture}[baseline=-.1cm]
    \draw        (0,0)   -- (1/4,0);
    \draw        (1,0) -- (5/4,0);
    \draw [->]   (1/4,0) arc (180:100:3/8);
    \draw [->]   (1,0) arc (0:80:3/8);
    \draw [->]   (1,0) arc (0:-80:3/8);
    \draw [->]   (1/4,0) arc (180:260:3/8);
    \filldraw [gray!50] (5/8,3/8) circle (2pt);
    \draw (5/8,3/8) circle (2pt);
    \filldraw [gray!50] (5/8,-3/8) circle (2pt);
    \draw (5/8,-3/8) circle (2pt);
\end{tikzpicture} \nonumber\\
    & = \int^>_{\hat{q}}  N_0 (\hat{k}/2 + \hat{q})
                          N_0 (\hat{k}/2 - \hat{q})
                          V_0 (\mathbf{k}, \mathbf{k}/2 + \mathbf{q})
                          V_0 (-\mathbf{k}, -\mathbf{k}/2 - \mathbf{q}) \,.
\end{align}%
\end{subequations}
In the expressions above, $V_0$ is defined in Eq.~\eqref{eq:bareVertex} in the main text, and  $G_0$ and  $N_0$ are given below Eq.~\eqref{eq:LangevinFourier} and below Eq.~\eqref{eq:bareVertex}, respectively.
Loop integrals are first computed over the frequencies using the residue theorem and then reducing the $d$-dimensional integral over the internal momentum $\mathbf{q}$ to a one-dimensional integral over its modulus $|\mathbf{q}| = q$.
%
The integration over the internal momentum $q$ is then facilitated by performing an expansion in series of $q\to\infty$ and by keeping only the ultraviolet leading-order singularity.
This procedure simplifies the computation while allowing the determination of the $1/\varepsilon$ poles (where $\varepsilon=d_c-d$ with $d_c=6$ is the upper critical dimension) encoding the critical scaling of the theory.
The integration over the modulus $q$ itself is then obtained as $\int^>_{q} f(q) = f(\Lambda) \Lambda \delta\ell + \mathcal{O}(\delta\ell^2)$. 

In order to properly define the one-loop corrections $\delta\mu_{1,2}$ to the non-linear terms $\mu_{1,2}$, it is convenient to evaluate $V_1 = V_1^{\rm (a)}+V_1^{\rm (b)}+V_1^{\rm (c)}$ at external momenta $\mathbf{k}$ and $\mathbf{k}_1$ such that $|\mathbf{k}_1|=|\mathbf{k}|=k$ and $\mathbf{k} \cdot \mathbf{k}_1 = k^2 \cos\theta$.
This determines the vertex function $V_1\equiv V_{1,\theta}$ solely in terms of $\theta$.
Here, we report the one-loop corrections to the various coefficients in the model~\eqref{eq:Langevin}; being interested in the RG flow near the critical point, all the corrections listed below are evaluated at $\sigma=0$ except for Eq.~\eqref{eq:oneloopmass} for $\sigma$ itself, in terms of which we determine the critical point given by Eq.~\eqref{eq:massShift}:
\begin{subequations}\label{eq:oneloopquantities}
\begin{align}
\delta\sigma & = \left. G_1(\hat k)\right|_{\hat k= \hat 0} \nonumber\\
& = - \frac{K_d \Lambda^{2-\varepsilon}\mathcal{D}_0}{D^2\tau}
  \left[2\lambda^2 - \left(\frac{3}{2} + \frac{1}{d}\right) \mu_1 \lambda  - \left(1-\frac{6}{d}\right) \mu_2\lambda   \right] \delta\ell +
  \frac{K_d \Lambda^{-\varepsilon}\mathcal{D}_0\sigma}{D^3\tau}
    \left[4\lambda^2 - \left(3 + \frac{3}{d}\right) \mu_1 \lambda  - \left(2-\frac{14}{d}\right) \mu_2\lambda   \right] \delta\ell \label{eq:oneloopmass} \, , \\
  \delta \tau & = \left. \ii \partial_{\omega_k} G_1(\hat k) \right|_{\hat k = 0} \nonumber\\
  & = K_d \Lambda^{-\varepsilon} \frac{\mathcal{D}_0}{D^3 \tau}\tau \left[ \lambda^2 - \left(\frac{3}{4} + \frac{1}{2d} \right) \mu_1\lambda - \left(\frac{1}{2}-\frac{3}{d} \right) \mu_2\lambda \right] \delta\ell  \,, \\
  \delta D & = \left. -\partial_{k^2} G_1(\hat k) \right|_{\hat k= \hat 0} \nonumber\\
  & =  \frac{K_d \Lambda^{-\varepsilon}\mathcal{D}_0}{D^3\tau} D \left[
  \frac{\mu_1\lambda}{4} \left(\frac{9}{d}-\frac{6}{d+2}-1\right) + \mu_2\lambda \left(\frac{3}{2}-\frac{23}{2d}+\frac{21}{d+2}\right)
  +\lambda^2 \frac{d-2}{d}
  + \frac{\mu_1^2}{4}\left(\frac{10}{d}-\frac{6}{d+2}-3 \right) + \right. \nonumber\\
  &\left.
  + \mu_1\mu_2\left(\frac{9}{d+2}-\frac{6}{d}-2 \right)
  + \mu_2^2\left(\frac{6}{d}-1 \right)
  \right] \delta\ell \,, \\
\delta \mu_1 & =  \delta\lambda - \frac{9}{16}V_{1,\theta=\pi}+\frac{25}{16} V_{1,\theta=\pi/2} \nonumber\\
    & = -\frac{K_d \Lambda^{-\varepsilon}\mathcal{D}_0}{D^3\tau} \lambda \left[
    \lambda (2\mu_2-\mu_1)
    + \mu_1^2 \left(\frac{3}{4}-\frac{3}{2d}+\frac{3}{d+2}\right)
    - \mu_1\mu_2 \left(1 -\frac{16}{d}+ \frac{28}{d+2} \right)
    - \mu_2^2 \left(1 +\frac{42}{d} -\frac{60}{d+2}  \right)
    \right] \delta\ell \, , \\
\delta \mu_2 & = \frac{5}{4}\delta\lambda + \frac{15}{32} V_{1,\theta=\pi} + \frac{25}{32} V_{1,\theta=\pi/2} \nonumber\\
& = \frac{K_d \Lambda^{-\varepsilon}\mathcal{D}_0}{D^3\tau} \lambda \left[
\mu_1^2  \left(\frac{1}{d}-\frac{3}{2(d+2)} \right)
- \mu_1\mu_2  \left(\frac{10}{d}-\frac{14}{d+2} \right)
+ \mu_2^2  \left(\frac{24}{d}-\frac{30}{d+2} \right)
\right] \delta\ell \,,  \\
\delta \lambda & = - \left. V_1(\hat k, \hat k/2+\hat k_1) \right|_{\hat k=\hat 0} \nonumber\\
& = \frac{K_d \Lambda^{-\varepsilon}\mathcal{D}_0}{D^3\tau} \lambda  \left[
4\lambda^2
-\mu_1\lambda \left(5+\frac{6}{d}\right)
- \mu_2\lambda \left(6-\frac{28}{d}\right)
+ \mu_1^2 \left(\frac{3}{2}+\frac{15}{2d}-\frac{9}{2(d+2)} \right)
+  \mu_1\mu_2 \left(4-\frac{48}{d}+\frac{42}{d+2} \right) \right. \nonumber\\
& \left.
+  \mu_2^2 \left( 2+\frac{58}{d}-\frac{90}{d+2} \right)
\right] \delta\ell \,,\\
\delta \mathcal{D}_0 & = \left.N_1(\hat k)\right|_{\hat k= \hat 0} \nonumber\\
& = \frac{K_d \Lambda^{-\varepsilon}\mathcal{D}_0}{D^3\tau} \mathcal{D}_0\lambda^2 \delta\ell \,.
\end{align}
\end{subequations}
Equations~\eqref{eq:RGfunctions} in the main text are finally obtained by evaluating Eq.~\eqref{eq:oneloopquantities} at the upper critical dimension $d = d_c = 6$ and after rescaling the variables as described in the main text.
Note that the fluctuation-induced shift of the critical value $\sigma_c$ of the parameter $\sigma$ discussed in the main text is obtained from Eq.~\eqref{eq:oneloopmass}. In terms of the rescaled variables, it reads:
\begin{align} \label{eq:massShift}
  \sigma_c = - \frac{\Lambda^2}{D}
    \left[2\bar\lambda^2 - \left(\frac{3}{2} + \frac{1}{d}\right) \bar\mu_1 \bar\lambda  - \left(1-\frac{6}{d}\right) \bar\mu_2\bar\lambda \right] \, .
\end{align}

\section{Extrapolated fixed points in spatial dimensions far from $d_c$}

Evaluating the corrections at $d=d_c$ is consistent with the rationale of the $\varepsilon$-expansion, which is based on a perturbative dimensional expansion valid close to $d_c$.
For completeness, however, we have reported the one-loop corrections in Eq.~\eqref{eq:oneloopquantities} with the $d$-dependent coefficients which result from evaluating the corresponding loop integrals. Any extrapolation of these coefficients to $d < d_c$ generally leads to uncontrolled, scheme-dependent predictions typical of ``fixed-$d$ RG'' computations~\cite{frey1994twoloop,lassig1995renormalization,wiese1998perturbation}. 

Nonetheless, it is informative to investigate the one-loop flows away from the upper critical dimension. To this aim, it is convenient to first simplify the one-loop RG flows by enforcing the exponent identity $z+\chi=0$ which follows from the pseudo-Galilean symmetry discussed in the main text. 
In addition to the line of fixed points in the plane $\lambda^*=0$ discussed in the main text, which are unstable in the $\lambda$ direction, we find the following five additional fixed points $(\mu_1^*,\mu_2^*,\lambda^*)$ of the RG flow equations:
\begin{enumerate}
    \item The Gaussian fixed point with $\mu_1^{*,0}=\mu_2^{*,0}=\lambda^{*,0}$ which is stable above the critical dimension $d_{\rm c}=6$ and unstable below.
    
    \item Two fixed points $\pm(\mu_1^{*,1},\mu_2^{*,1},\lambda^{*,1})$ with (here and below we report the expressions of the dimensionless variables, but we drop the bar on the corresponding symbols to lighten the notation)
\begin{align} \label{eq:fixed-set1}
    \lambda^{*,1} = \frac{5}{6} \mu_1^{*,1} = 5 \mu_2^{*,1}
    = \sqrt{\frac{5 d (d-6)(d+2)}{2[7d^2 - 10d -12]}},
\end{align}    
which are real-valued for $d< \frac{5+\sqrt{109}}{7}\approx 2.21$ or $d \geq 6$. These fixed points are only stable for $d<2$ and unstable otherwise.    

    \item Two fixed points $\pm(\mu_1^{*,2},\mu_2^{*,2},\lambda^{*,2})$ with 
\begin{align}   \label{eq:fixed-set2}
   \lambda^{*,2} = \left[1+\frac{2}{d(d-8)} \right] \mu_1^{*,2}  
   = 2 \left[ 1 + \frac{2(1-2d)}{d(d-4)} \right] \mu_2^{*,2} 
   = \sqrt{\frac{2(d-6)(d+2)(d^2-8d+2)^2}{5d^6-97d^5+606d^4-1358d^3+1320d^2-872d+96}} \,.
\end{align}
These fixed points are real-valued for $ 0.13 \lesssim d\lesssim 2.3$, and also for $6 \leq d \lesssim 7.54$ or $d \gtrsim 8.55$ above the upper critical dimension. They are stable RG fixed points only for $ 2 \leq d \lesssim 2.3$.
\end{enumerate}

\begin{figure*}[h!]
	\begin{minipage}[c]{.32\linewidth}
		\centering
		\includegraphics[width=\linewidth]{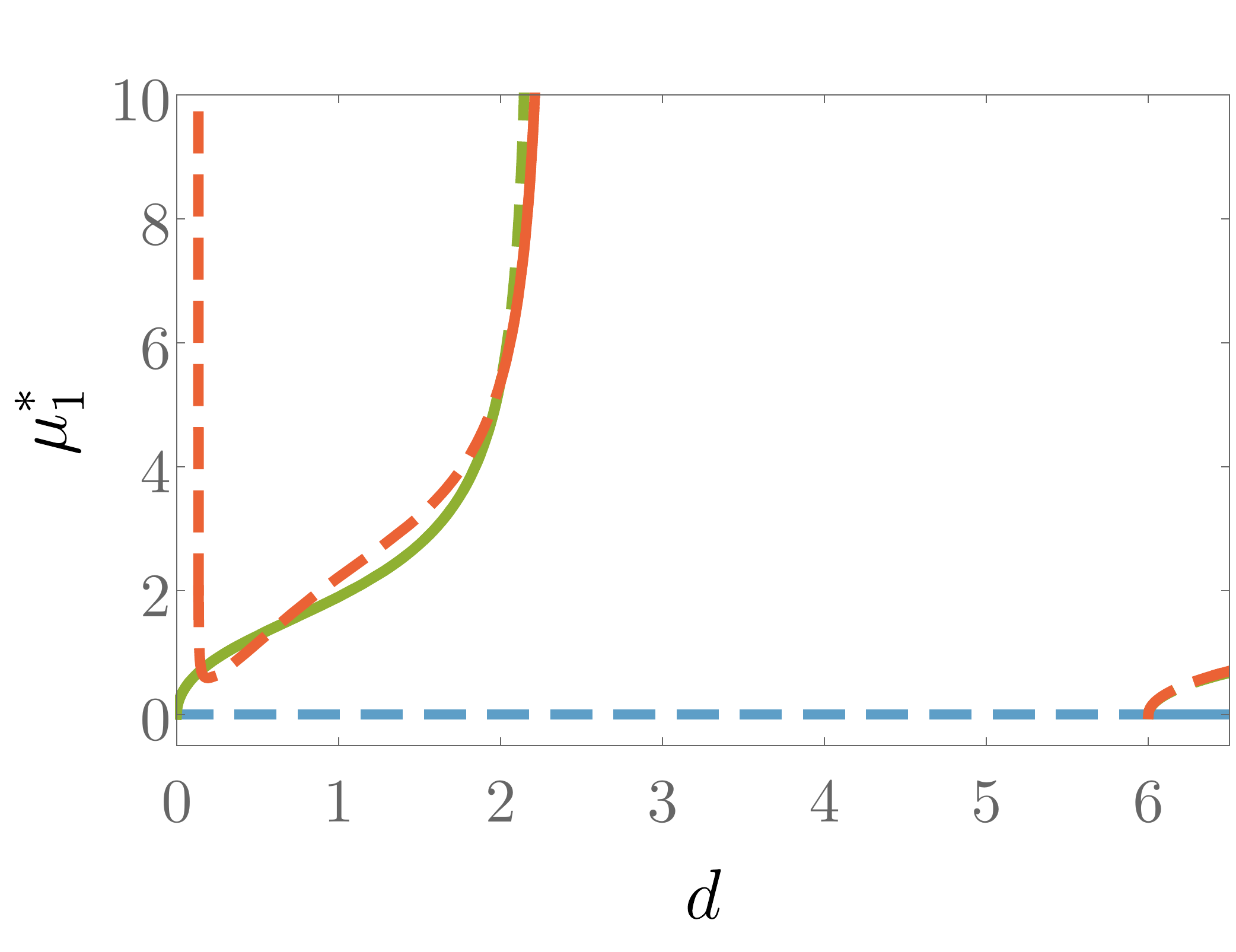}
	\end{minipage}  
    \hfill
	\begin{minipage}[c]{.32\linewidth}
		\centering
        \includegraphics[width=\linewidth]{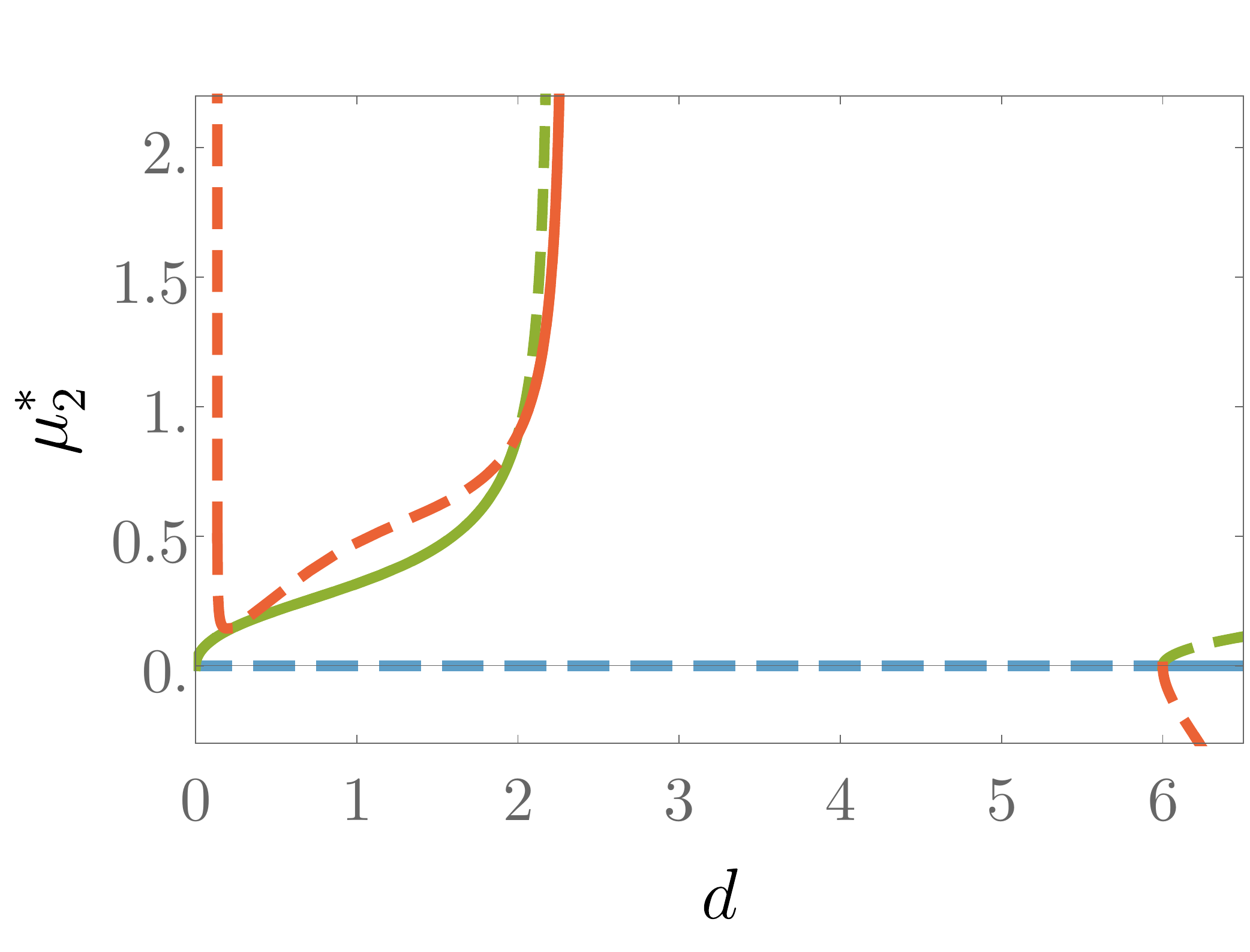}
	\end{minipage} 
    \hfill
	\begin{minipage}[c]{.32\linewidth}
		\centering
		\includegraphics[width=\linewidth]{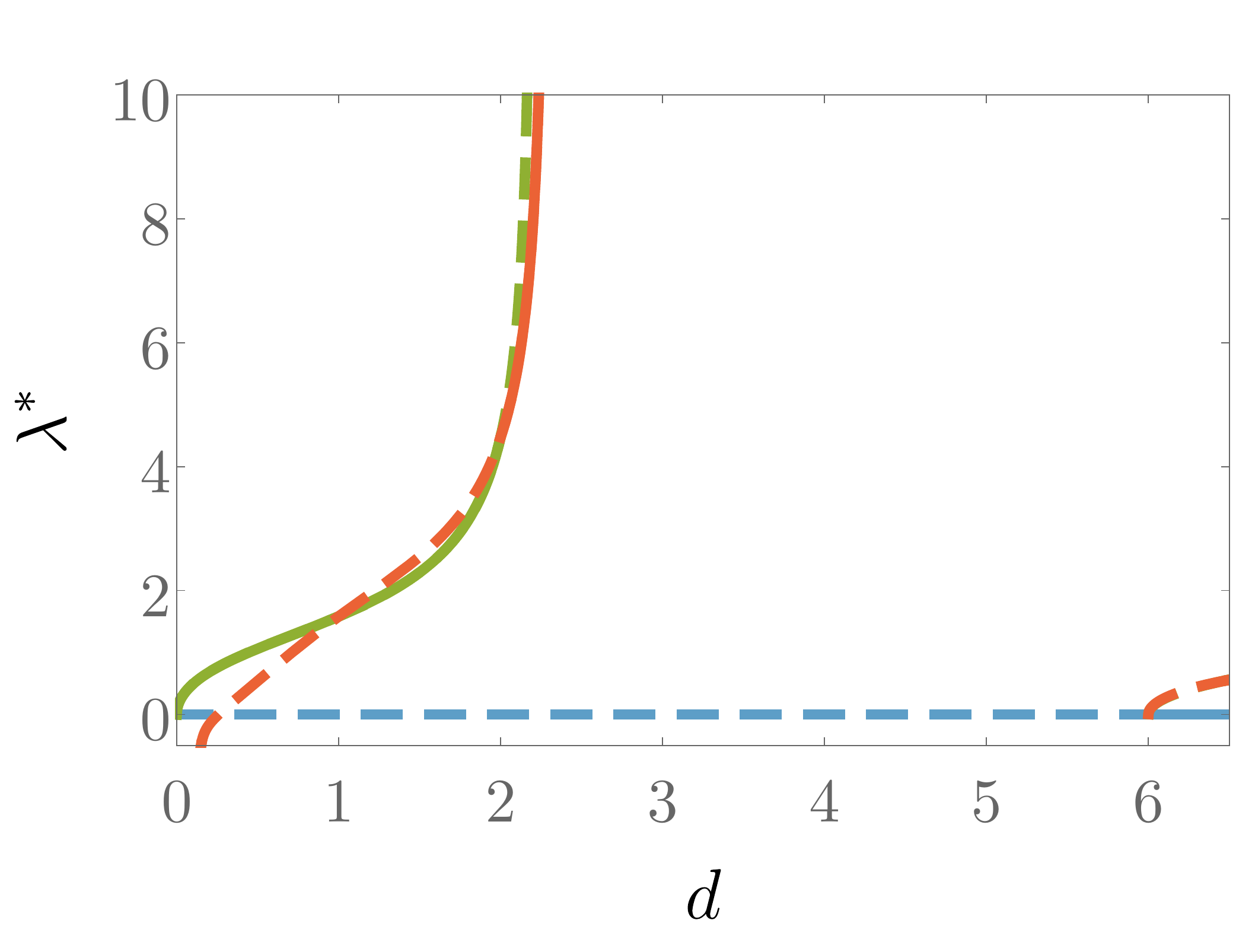}
	\end{minipage} 
	\caption{Coordinates $\mu_1^*$ (left), $\mu_2^*$ (middle) and $\lambda^*$ (right) of the RG fixed points as functions of the spatial dimension $d$. 
	Solid lines refer to the stable fixed points and dashed lines to unstable ones.
	The Gaussian fixed point is plotted in blue, the fixed point $(\mu_1^{*,1},\mu_2^{*,1},\lambda^{*,1})$ given by Eq.~\eqref{eq:fixed-set1} is shown in green, and the fixed point $(\mu_1^{*,2},\mu_2^{*,2},\lambda^{*,2})$ given by Eq.~\eqref{eq:fixed-set2} is shown in red.
} \label{fig:coordinates}
\end{figure*}

In Fig.~\ref{fig:coordinates} we summarize this analysis by plotting the three coordinates $(\mu_1^*,\mu_2^*,\lambda^*)$ of the fixed points discussed above as functions of the spatial dimension $d$. At one-loop, our analysis reveals no stable fixed point of the RG flow equations in $2.3\lesssim d<6$. A stable fixed emerges at $d\simeq 2.3$ (orange solid line and green solid line) and can be followed up to $d=0$. Determining whether this fixed point captures correctly the physics in $d=1$ and 2 or whether it is an artefact of the one-loop computation is beyond the scope of this work.

\twocolumngrid

\bibliographystyle{apsrev4-1}
\bibliography{biblio-EPL.bib}

\end{document}